\documentclass{ws-procs9x6}

\usepackage{amsmath,amssymb}
\usepackage{graphicx}
\usepackage{cite}

\begin{document}

\begin{center}
{\Large\textbf{de Sitter Duality and Holographic Renormalization}}\end{center}

\begin{flushright}
KEK-TH-2382
\end{flushright}

\begin{center}
Yoshihisa \textsc{Kitazawa}$^{1),2),}$
\footnote{E-mail address: kitazawa@post.kek.jp} 
\end{center}

\begin{center}
$^{1)}$
\textit{KEK Theory Center, Tsukuba, Ibaraki 305-0801, Japan}\\
$^{2)}$
\textit{Department of Particle and Nuclear Physics}\\
\textit{The Graduate University for Advanced Studies (Sokendai)}\\
\textit{Tsukuba, Ibaraki 305-0801, Japan}\end{center}

\begin{abstract}
 Applying the renormalization group, we derive the stochastic equations as the effective theory at the horizon. We focus on the conformal zero mode to respect local Lorentz symmetry.
 Under Gaussian approximation, we derive the  fundamental  equation for the Universe (EqU).
 We  also derive the identical equation from the first law of thermodynamics in a dual geometric picture. 
There is a convincing  evidence for de Sitter duality between quantum stochastic physics on the boundary and classical thermodynamics in the bulk.
The equation for the Universe (EqU)
possesses the solution with the ultraviolet fixed point.
It also contains the inflationary universe with the power potentials.
We discuss possible scenarios for the very early universe with decreasing $\epsilon$.
We argue inflationary universe subsequently dominates  to maximize the entropy and $\epsilon$ problem is naturally solved.
\end{abstract}


\bodymatter

\section{Introduction}

The importance of stochastic effect has been long recognized  in inflation theory\cite{Starobinsky1986, Tsamis2005}.
We have recognized that inflaton may very well be conformal zero mode
and Einstein gravity and inflation theory are dual to each other.\cite{Kitamoto2014}.
In this work we demonstrate the consistency and consequences )of  de Sitter duality.

dS space may be decomposed into the bulk and the boundary, i.e., the sub-horizon and horizon. 
From a holographic perspective, we consider the conformal zero mode dependence of the Einstein-Hilbert action: 
\begin{align}
\frac{1}{16\pi G_N }\int \sqrt{g}d^4x(Re^{2\omega}-6H^2e^{4\omega}) 
\simeq\frac{\pi}{G_N H^2}(1 -4\omega^2).
\label{action1}\end{align} 
Our gauge fixing procedure and the propagators are explained in \cite{Kitamoto2012}.
It does not contribute to the interaction vertices. 
The semi-classical dS entropy was obtained 
by rotating the background spacetime $dS^4$ to $S^4$ in (\ref{action1})
\cite{Gibbons1977}. 

The quadratic part of $\omega$ constitutes a Gaussian distribution function for the conformal zero mode, 
\begin{align}
\rho(\omega)=\sqrt{\frac{4}{\pi g}}\exp\big(-\frac{4}{g}\omega^2\big).
\label{distribution1}\end{align}
It may represent an initial state of the Universe when the dS expansion begins. 
In order to describe the time evolution of the conformal factor of the Universe, 
we introduce a new parameter $\xi(t)$.
\begin{align}
\rho(\xi (t),\omega)=\sqrt{\frac{4\xi (t)}{\pi g}}\exp\big(-\frac{4\xi (t)}{g}\omega^2\big). 
\label{distribution2}\end{align}
$\xi$ is the only parameter in the Gaussian approximation.
We work within the Gaussian approximation since it is an excellent approximation 
for gravity with the small coupling $g$.
The von Neumann entropy $S=-\text{tr}(\rho \log \rho)\sim {1/2\log (g/\xi )}$ becomes larger as $\xi$ becomes smaller. Thus the diffusion triggers an instability in de Sitter space.

In terms of  the distribution function, the $n$-point functions are defined as follows 
\begin{align}
\langle\omega^n(t)\rangle _{\text{boundary}} =\int d\omega \rho(\xi(t),\omega)\omega^n. 
\label{n-pt}\end{align}
In particular, the two-point function of the conformal mode is given by 
\begin{align}
\langle\omega^2(t)\rangle_{\text{boundary}}=\frac{g}{8\xi(t)}.
\label{w2}
\end{align}
$1/\xi$ is the enhancement factor of the conformal (scalar) perturbation over tensor mode.

The negative norm of the bulk conformal mode indicates that 
the $\rho$ diffuses toward the future. 
In fact the perturbative quantum expectation of (\ref{action1}) gives  $g(t)\sim g(1-2\gamma Ht)\sim g(1-3gHt)$. 
The duality between the inflation  and quantum 
gravity is also based on this one loop effect \cite{Kitamoto2019-1,Kitamoto2019-2}.
However such an estimate is reliable only locally $Ht\ll 1$. In cosmology it is essential to resum all powers of IR logarithms $(Ht)^n$ to understand the global picture. For such a purpose, we find our holographic approach is up to the task.
We investigate the dynamics of conformal mode after integrating the bulk mode.
The two point function gives rise to IR logarithms.
\begin{align}
<\omega^2>_{\text{bulk}}&=-\frac{3g}{4}\int_{Ha(t)}^{\Lambda }\frac{dk}{k} 
\notag\\
=&\frac{3g}{4}Ht=  \frac{3g}{4}N(t).\label{IRlog}\end{align}

We assume there is no time dependent UV contributions. 
We focus on the Hubble scale physics where $a(t)=1/-\tau H=\exp(Ht)$.
We recall that $-\tau\mu \sim 1$ at the Horizon.
While the wave functions of the bulk mode oscillates,
those of the boundary mode do not. That is why we call  it the zero mode.
Our strategy is to integrate out oscillating mode first.
The finite bare distribution function is given by subtracting the bulk mode contribution above the Hubble scale. We thus construct low energy effective theory around the Hubble scale. Such a theory is holographic and finite after the subtraction.
\begin{align}
\rho_B=\exp\big(-\frac{3g}{4}\frac{Ht}{2}\frac{\partial^2}{\partial\omega^2}\big)\rho .
\label{bare}\end{align}

The renormalization scale of the low energy effective action is the Hubble scale.
As $\rho_B$ is independent of the renormalization scale,
the renormalized distribution function obeys the following renormalization group equation. 
\begin{align}
\dot{\rho}
-\frac{3g}{4}\cdot\frac{H}{2}\frac{\partial^2}{\partial \omega^2}\rho
=0, 
\label{FP0}\end{align}
where $\dot{O}$ denotes a derivative of $O$ with respect to the cosmic time $t$. 
The factor $3g/4$ in the diffusion term is the projection factor to the IR region. 
The conformal mode $\omega$ consists of a minimally coupled field $X$. 
$\omega=\sqrt{3}X/4+Y/4$ \cite{Tsamis1994,Kitamoto2012}. 
We neglect $Y$ because it has the effective mass $m^2=2H^2$. 
There is no drift term in the reduced space consisting only of $X$. 

The gravitational FP equation (\ref{FP0}) is  obtained by
integrating  the  quantum bulk modes inside the horizon. It turns out to be a diffusion equation due to the lack of the drift term.
The solution is the Brownian motion as it is jolted by the horizon exiting modes.
The FP equation is a dynamical renormalization group equation. 
We can sum up the IR logarithms $\log ^na=(Ht)^n$ by this equation to find a running coupling $g(t)$. 
The FP (diffusion) equation 
shows that the solution is the Gaussian distribution with the  standard deviation 
increasing linearly with the e-folding number $N(t)$ \cite{Parisi}.
\begin{align}
{3\over 4}gHt={3g\over 4}N(t)={g\over 8\xi}.
\label{xi}\end{align}
The standard deviation is related to $N$ as
$1/\xi=6N$ in (\ref{xi}).
It is consistent with a standard Brownian motion prediction.
The von Neumann entropy thus increases logarithmically, 
\begin{align}
\delta S=\frac{1}{2}\log {1\over\xi}
=\frac{1}{2}\log 6N(t). 
\label{entropy1}\end{align}

Identifying the von Neumann entropy of conformal zero mode
with the quantum correction to dS entropy,
we obtain the bare action with the counter term
\begin{align}
\frac{1}{g_B}=\frac{1}{g(N)} - \frac{1}{2}\log(6N). 
\label{coupling}\end{align}
By requiring the bare action is  
independent of the renormalization scale: namely  $N$ ,
we obtain the one loop $\beta$ function.
\begin{align}
\beta=\frac{\partial}{\partial \log(N)}g(N)=-\frac{1}{2}g(N)^2. 
\label{beta1}\end{align}
We find the running gravitational coupling as
\footnote{The coefficient in front of $N$ can be put to the identity.}
\begin{align}
 g(N)={2\over \log(N)}. 
\label{beta1}\end{align}

The holographic investigation at the boundary shows that 
$g$ is asymptotically free toward the future \cite{Gross1973,Politzer1973}. 
The renormalization group trajectory must reach Einstein gravity in the weak coupling limit 
for the consistency with general covariance \cite{Kawai1993}. 
We find that it approaches a flat spacetime in agreement with this requirement.

\section{Equation for the Universe and de Sitter Duality}
The Gaussian distribution of the conformal zero mode is characterized by the standard deviation $1/\xi$.  Although there is no inflaton in Einstein gravity, 
we propose to identify the inflaton $f^2$ as $f^2\propto 1/\xi$. 
In our interpretation, the inflaton is not a fundamental field but a 
stochastic variale. 
The two point function at an equal time grows due to the Brownian motion:  IR logarithmic fluctuations $1/\xi\sim N $.
While the inflation theory is specified by the inflaton potential, 
the dynamics of quantum gravity is determined by the FP equation 
which describes the stochastic process at the horizon. 
We thus argue 
the classical solution of the inflation theory satisfies the FP equation as well. 

It is likely that there are multiple elements in the universality class of quantum gravity/inflation theory. 
The inflation era of the early Universe may be one of them. As we discuss shortly
we find a pre-inflation era which  is indispensable to launch inflation era
which in turn necessary to trigger the big bang.
We evaluated the time evolution of entropy to the leading log order in (\ref{entropy1}). 
In order to take account of the higher loop corrections in $g$, the FP equation 
should be generalized. It turns out to be just necessary to make the dimensionless gravitational coupling time dependent $g(t)$:
\begin{align}
{\partial\over \partial t}{\rho}
-\frac{3g(t)}{4}\cdot\frac{H}{2}\frac{\partial^2}{\partial \omega^2}\rho
=0.
\end{align}
The Gaussian ansatz (\ref{distribution2}) is an exact solution if  $g$ is 
a constant. 

In what follows, we consider such a solutions.
\begin{align}
\rho=\sqrt{\frac{4\xi (t)}{\pi g(t)}}\exp\big(-\frac{4\xi (t)}{g(t)}\omega^2\big).
\end{align}
We next put the ansatz into the FP equation and
find the condition for the background to satisfy.
In the small $\omega$ limit, 
the background must satisfy
\begin{align}
{\partial \over \partial t}
{1\over 2}\log({\xi\over g(t)})+3H\xi=0.
\end{align} 

We obtain 	a remarkably simple equation in terms of $N=Ht$.
\begin{align}
\frac{\partial}{\partial N}\log \frac{g(t)}{\xi}=6\xi. 
\label{FP3}\end{align}
(\ref{FP3}) determines the evolution of von Neumann entropy $S={1\over 2}\log{g\over \xi}$ with respect to $N$.
This formula shows the validity of our postulate that von Neumann entropy
of conformal zero mode constitutes the quantum correction to de Sitter entropy.
We call it the equation for the Universe (EqU).

The equation (\ref{FP3}) has inflationary solutions with power potentials.
\begin{align}
g=c\tilde{N}^\frac{m}{2},\hspace{1em}\xi=\frac{m+2}{12\tilde{N}}.
\label{solution2}\end{align}
Here $c$ is an integration constant.
$m$ denotes the power of the potential: $f^m$.
It is convenient to 
replace $N$ by $\tilde{N}$
where $\tilde{N}=N_e-N$.  $N_e$ denotes the e-foldings at the end of inflation. 
 (\ref{FP3}) becomes as follows after the change of the variable,
\begin{align}
-\frac{\partial}{\partial N}\log \frac{g(\tilde{N})}{\xi(\tilde{N})}=6\xi(\tilde{N}). 
\label{FPtr}\end{align}

Although the dS entropy can be explained by quantum effects alone for the weakly coupled inflaton solution, 
the strongly coupled inflaton solution is a dual object in the sense that geometrical description is reliable. 
The increase of the entropy $S=1/g$ can be evaluated by the first law $T\Delta S=\Delta E$ 
where $\Delta E$ is the incoming energy flux of the inflaton  \cite{Frolov2003}.
In this way, the one of the Einstein's equation is obtained:
\begin{align}
\dot{H}(t)=-4\pi G_N \dot{f}^2.
\end{align}
From this formula, we obtain 
\begin{align}
2\epsilon=-{\partial \over \partial {N}}\log g(t).
\label{MEsv}
\end{align}
We add the same quantity ${1/ \tilde{N}}$ to the both sides of the equation.
\begin{align}
2\epsilon+{1\over \tilde{N}}=-{\partial \over \partial {N}}\log (g \tilde{N}).
\label{mf2}
\end{align}

For power potential inflationary universe, (\ref{mf2}) is rewritten as
\begin{align}
6\xi={\partial \over \partial {N}}\log {g\over \xi}.
\end{align}
It is precisely our EqU (\ref{FP3}).
We have derived it in a dual geometric picture here in place of the original quantum stochastic picture. 
It proves the consistency of the EqU and the confirmation of de Sitter duality.
Since (\ref{FP3}) and (\ref{MEsv}) are equivalent, the solutions of the former
satisfy the latter.

\section{Inflation and UV completion}

In the literature, $\delta N$ formalism is widely used to investigate the curvature perturbation. It underscores the validity of the stochastic picture of the inflation.
\cite{Starobinsky1985,Bond,Stewart,Lyth2005a,Lyth2005b}
 Let us consider the fluctuation of the curvature perturbation $\zeta$.
\begin{align}
&{ \zeta} = {\delta N} ={H\over \dot{ f}}\delta f ,\notag\\
&<\delta f(t) \delta f(t')> =({H^2\over 4\pi^2})H\delta(t-t').
\label{Lgdn}
\end{align}
We obtain in the super-horizon regime:                                                                                                                 
\begin{align}
&<\zeta^2(t)> =<  ({H\over \dot{f}})^2\delta f^2>\notag\\
&={1\over 2\epsilon M^2_P}<\delta f^2>\notag\\
&={H^2N(t)\over 8\pi^2\epsilon M^2_P}={g\over \epsilon(t)}=P.
\label{P123}
\end{align}

We recall the following identity holds at the horizon exit $t=t_*$
\footnote{We follow the notations of \cite{Maldacena203}.}
\begin{align}
\dot{\rho}(t_*)e^{\rho(t_*)}=k.
\end{align}
 It is nothing but choosing our renormalization scale as $\log k = Ht$.
 Let $P\sim k^{1-n_s}$. The scaling dimension of $P$ can be estimate as
 \begin{align} 
& k{d\over dk}(4\log \dot{\rho}(t_*)
 -2 \log\dot{\phi}(t_*))\notag\\
 =&{d\over dN_*}
(4\log \dot{\rho}(t_*)
 -2 \log\dot{\phi}(t_*)) \notag\\
=&2(\eta-3\epsilon),
\end{align}
where
\begin{align}
\epsilon={\dot{\phi}^2\over 2 H^2},~~
\eta={\ddot{\phi}\over \dot{\rho}\dot{\phi}}+\epsilon.
 \end{align}

In power inflation potential models $V\sim f^m$, 
$\xi$ and $\epsilon$ scale as $1/N$
due to the universality of the random walk. 
They belong to the same universality class as the following
scalar two point functions scale as
\begin{align}
<\zeta^2> ={g\over \epsilon}=4 g N(t),~~ \epsilon ={m\over 4N(t)}.                                
\end{align}
The scaling exponents agree with our FP equation and the $\delta$ N formalism.
\begin{align}
1-n_s={\partial \over \partial N}\log (g N(t))=2\epsilon+{1\over N(t)}={m+2\over  2N(t)}.
\end{align}
The conformal mode is indistinguishable from the curvature perturbation as far  as the scaling property is concerned. 
Our identification of the conformal mode and the inflaton therefore works in generic  quantum gravity with conformal  mode or inflation theory. The advantage of conformal mode is  being Lorentz scalar, well understood in covariant field theory. The precise normalization of $P=g/\epsilon$ needs to be fit with the data: - normalization.
It is still a challenge to separate $g$ and $\epsilon$.
 
There is a UV fixed point in our renormalization group. We study it next.
FP equation (\ref{FP3}) enables us to evaluate higher order corrections to the $\beta$ function. 
The expansion parameter is $1/\log N$. 
We can confirm that the following $g_f$ and $\xi_f$ satisfies (\ref{FP3}), 
\begin{align}
g_f=\frac{2}{\log N}\big(1-\frac{1}{\log N}\big),\hspace{1em}
\xi_f=\frac{1}{6N}\big(1-\frac{1}{\log N}\big). 
\label{solution1}\end{align}
Thus, the $\beta$ function, $\epsilon$ and the semi-classical entropy generation rate are given by 
\begin{align}
\beta=\frac{\partial}{\partial \log N}g_f=
 -{2\over \log^2 N}+{4\over \log(N)^3}= 
 - 2({1-2\frac{1}{\log N}}){1\over \log^2 N}.
\label{beta2}\end{align}

\begin{align}
\epsilon_f = -{1\over 2}{\partial \over \partial N}\log(g_f)=
 -{1\over 2g_fN}\beta_f.
\label{ep3}
\end{align}

\begin{align}
\frac{\partial}{\partial N}S_{sc}=\frac{\partial}{\partial N}\frac{1}{g_f}
=-\frac{1}{Ng_f^2}\beta_f.
\label{entropy2}\end{align}

Before concluding this section, we explicitly construct  the logarithmic solution of
(\ref{FP3}). We start with the following basic structure and decorate on it. 
\begin{align}
{\partial \over \partial N}\log({2N})
 ={1\over N}.
 \label{ep4}\end{align}
Let us subtract  $({1}/{2})({\partial\over  \partial N}\log(\log N))={1}/{2(N\log N)}$ on the both sides
of (\ref{ep4}).
\begin{align}
{\partial \over \partial N}\log(\frac{2N}{\log N})
 ={1\over 2N}-{1\over 2N\log N}.
\end{align}
This equation is equivalent to (\ref{FP3}).

A remarkable feature is that the coupling has the maximum value $g=1/2$ at the beginning. 
It steadily decreases toward the future 
as the $\beta$ function is negative in the whole region of time flow. 
It has two fixed points at the beginning and at the future of the Universe. 
The existence of the UV fixed point may indicate the consistency of quantum gravity. 
The single stone solves the
$\epsilon$ problem \cite{Penrose1988} as well since it vanishes at the fixed point. 
The $\beta$ function describes a scenario that 
our Universe started the dS expansion with a minimal entropy $S=2$ while it has $S=10^{120}$ now. It corresponds to $N=e^2\sim 7.4,  a=e^N\sim 1.6 \times 10^3$. 
The just born Universe is rather large which reflects the critical  coupling $g=1/2$ is rather small. In terms of the reduced Planck mass, $H^2/{4\pi}^2 M_P^2=1$. 

Since we work with the Gaussian approximation, our results on the UV  fixed points are not water tight
as the coupling is not weak. Nevertheless we find it remarkable that they
support the idea that quantum gravity has a UV fixed point 
with a finite coupling. In fact 4 dimensional de Sitter space is constructed in the target space 
at the UV fixed point of $2+\epsilon$ dimensional quantum gravity  \cite{Kawai1993}.
4 dimensional de Sitter space also appears at the UV fixed point of the exact
renormalization group \cite{Reuter}\cite{Souma}.

Such a theory might be a strongly interacting conformal field theory. 
However, it is not an ordinary field theory as the Hubble scale is Planck scale. 
Our dynamical $\beta$ function is closely related to the cosmological horizon and physics around it. 
The existence of the UV fixed point could solve the trans-Planckian physics problem. 
A consistent quantum gravity theory can be constructed 
under the assumption that there are no degrees of freedom at trans-Planckian physics \cite{Bedroya2019}. 
In this sense, it is consistent with string theory and matrix models.  
The Universe might be governed by (\ref{solution1}) in the beginning 
as it might be indispensable to construct the UV finite solutions of the FP equation. 

\section{The Inflationary Universe and Pre-History}

The inflaton may be identified with the stochastic variable $f$ whose correlators show characteristic features of Brownian motion.  $<f^2>=\tilde{N}$ and $g\propto \tilde{N}^{m/2}=<f^m>$. 
$6\xi=1-n_s$ measures the extra tilt of the scalar two point function $k^{1-n_s}$ with respect to $k$. 

We thus conclude:
\begin{align}
&g=c\tilde{N}^\frac{m}{2},~~
\epsilon = -{1\over 2}{\partial \over \partial N}log(g)=\frac{m}{4\tilde{N}},\notag\\ 
&1-n_s=6\xi=\frac{m+2}{2\tilde{N}}.
\label{infsol1}\end{align} 

In this case, the concave power solutions can be obtained by formally replacing $m$ by $1/n$. We admit it is not entirely clear why concave potentials are relevant.  It is possible that
the convex potentials are already excluded by observations.

In contrast, there is more room for concave potentials to
accommodate the observational information with the judicious choice of $n$.
\begin{align}
&\epsilon =-{1\over 2}{\partial \over \partial N}log(g)=\frac{1}{4n\tilde{N}} ,\notag\\
&1-n_s=\frac{\frac{1}{n}+2}{2\tilde{N}}.
\label{infsol2}
\end{align}
In our view the inflaton behaves as an particle in the Brownian motion.
Its trajectory is made of infra-red fluctuations.
Nevertheless the convex and the concave potentials in the inflation models seems to form two distinct groups.

The comparison (\ref{solution1}) and (\ref{infsol2}) indicates the 
effective couplings are $1/\log {N}$ in the former
\begin{align}
\epsilon={1\over \log N\cdot 2N} {(1-{2\over \log N})\over (1-{1\over \log N})}.
\label{infsol3}
\end{align}
and $1/2n$ in the latter . 
\begin{align}
\epsilon={1\over 2n \cdot 2\tilde{N}} .
\end{align} 
So concave potentials are in the weak coupling regime while convex
potentials belong to the strong coupling regime.
We believe this corrependence should be taken seriously.
At the UV fixed point $\log N=2$ suggests $n=1$. Although
the $1/\log N$ correction becomes small only when $n\sim 3$, it is still close to the critical point.  
The concave potentials are promising avenue to explore right now \cite{2018}.


\begin{table}

\tbl{}

\centering

\begin{tabular}{ccccc}
&
& $n=1$
& $n=2$
&$n=3$
\\[.5pc] \hline\hline
&$\epsilon $
&${1\over 4\tilde{N}}$
& ${1\over 8\tilde{N}}$
& ${1\over 12\tilde{N}}$
\\[.5pc] \hline
&$1-n_s$
&$\frac{3}{2\tilde{N}}$
&$\frac{5}{4\tilde{N}}$
& $\frac{9}{8\tilde{N}}$
\\[.5pc] \hline

\end{tabular})

\label{tab:tm1} 

\end{table}


In Table \ref{tab:tm1}, we list expected $\epsilon$ and $1-n_s$ in the power potential
model with $f^{1/n},n=1,2,3$.
We note $1-n_s$ is bounded from below by $1/\tilde{N}$ while
$r=16\epsilon $ is not. It is consistent with the current observations.
It is important to establish the bound on $n$. We argue the order of magnitude
estimate can be trusted for the concave potentials as they are in the weak coupling regime.
The semi-classical de Sitter entropy favors smaller $n$ as the corresponding entropy grows faster: $S\sim (1/\tilde{N})^{1\over 2n} $.

Although it is still highly speculative, we argue that our Universe is likely to be located close to the point $m=n=1$.
Since the convex potentials are excluded by observation, it must be a concave potential with a small $n$.
The convex potentials correspond to strongly coupled systems. 
Here the naturalness and the anthropic principle may come in.
It is hard to believe that our Universe 
comes out of a strongly coupled system. 
It is much easier to accommodate hierarchies in the weakly coupled system. 
Our Universe might to be located near the boundary between the stable and unstable universes like the standard model of particle physics.
Since the Universe can stay at the fixed point forever, it is likely to be not far from it.

We recall here the curvature perturbation:
\begin{align}
P\sim {H^2\over (2\pi)^2 2M_P^2\epsilon}\sim 2.2 \times 10^{-9}.
\label{CP}
\end{align}
$P=<\zeta^2>=g/\epsilon$ must be generated during inflation.
It is bounded from below $g>10^{-11}$ if $\epsilon>1/200$.
In order to explain this magnitude, we must assume the inflation
lasted much beyond what CMB isotropy generation requires
as we examine it in the last part of this section.

The left-hand side of (\ref{FP3}) can be identified with $1-n_s$ 
where $1-n_s$ is the scalar spectral index. 
Let us recall that $1-n_s$ is expressed by the slow-roll parameters $\epsilon$ and $\eta$, 
\begin{align}
1-n_s=6\epsilon-2\eta. 
\label{relation}\end{align}
With the identification (\ref{infsol2}), the equation (\ref{FP3}) is equivalent to (\ref{relation}) 
for the $f^m$ inflaton potential where $\eta=(m-1)/(2\tilde{N})$. 
Our FP equation is expected to go beyond the resummation of leading IR logarithms which results in the logarithmic decay of $g$.
It is still a great surprise to find inflation theory as its solutions. 
Fortunately EqU can be derived in a dual geometrical picture.
This fact gives a non-perturbative evidence for the dS duality. 

The solution (\ref{solution1}) is UV complete. $g$ is attracted to the fixed point as we roll back the history of the Universe. 
However, it cannot terminate the eternal inflation 
as $\epsilon=-(1/2)\partial\log g /\partial N\sim 1/(2N\log N)$ decreases with time. 
On the other hand, the solution (\ref{solution2}) is not UV complete 
but it can end the inflation as $\epsilon\sim m/4\tilde{N}$ increases with time.   
These solutions generate the entropy in different ways: 
We consider the leading de Sitter entropy $1/g(t)$. The $t$ dependence of $g(t)$ depends on the 
quantum corrections through  the EqU equation.
$1/g\sim \log N$ for the former and $1/g\sim 1/\tilde{N}^\frac{1}{2n}$ for the latter. 
From the perspective of the dominant entropy principle, 
(\ref{solution1}) is chosen initially and (\ref{solution2}) is chosen after $\log N\sim 1/\tilde{N}^{1\over 2n}$. 
That is to say, $g_1$ in (\ref{solution1}) describes the newly born Universe and $g_2$ in (\ref{solution2}) describes the inflation era. 

The UV fixed point endows with a special initial condition for the inflation theory.  $\epsilon$ vanishes at the UV fixed point while it grows subsequently (\ref{infsol3}) up to $1/200$ at $N\sim 10$. It then decays again like $1/(2N\log N)$.
Let us investigate pre-inflation scenario.
The entropy grows logarithmicaly $1/g=(\log N)/2$ as $N$ grows after the Universe emerges out of the UV fixed point.
In the inflationary phase, the entropy grows power like $1/g\sim 1/\tilde{N}^{1/2n}$.
The entropy increases slowly  in the pre-inflation era while
it changes much faster in the inflation era. 
Inflation phase inevitably takes over the pre-inflation phase
as the entropy of the former dominates in time.

We may follow the evolution of $\epsilon$. It decreases as $1/(2N\log N)$ in the pre-inflation era and it
grows as $1/4\tilde{N}$ in the inflationary era . To the leading order of $N$, $\epsilon$ decay is canceled out by the subsequent growth during inflation. More precise scenario is $N=30,~\epsilon =1/200$ in the first stage. This reduction of $\epsilon$ will be  compensated by the growth in the second stage with $\tilde{N}=50$.  This is the minimum consistent scenario. However  $g$ is reduced  by only the factor $20$ in this process and we need to stretch the length of the both stages 
enormously: $10^{16} N\sim 10^{17}$ and $\tilde{N}=50 \rightarrow 10^{16}\tilde{N}$
by the common factor $10^{17} $ to explain $g\sim 10^{-10}$. 
$\epsilon$ decreases as small as $10^{-19}$ before bouncing back to $0(1)$.

During the inflation, 
the gravitational coupling decreases $\tilde{N}^{1\over 2n}$.
While $\epsilon$ decreases in the pre-inflation, it increases during the inflation.
 It is in the same ball park with the curvature perturbation $P\sim 10^{-9}$. 
 It is because $P=g/\epsilon$ and the inflation terminates at $\epsilon \sim 1$. We believe this coincidence is not an accident but an evidence for the existence of pre-inflation era prior to the conventional inflation era.
 
 The history of the Universe is likely to be much more versatile than the minimal inflation  which CMB
anisotropy requires. Inflation requires fine tuning such as minuscule $\epsilon$. 
We believe we have found a natural mechanism to accomplish it. 
So far in this section, we investigated pre-inflation to inflation transition as if it is the second order phase transition. There maybe a different picture. It may be possible to construct the
UV finite inflationary Universe as composite models \cite{Jain2007}.
Although it is beyond the scope of the paper, we have initiated some research along these ideas
\cite{HGRG}.

\section*{acknowledgments}
This work is supported by  Grant-in-Aid for Scientific Research (C) No. 16K05336. 
I thank Chong-Sun Chu, Masashi Hazumi, Satoshi Iso, Hikaru Kawai, Kozunori Kohri, 
Takahiko Matsubara, Jun Nishimura, Hirotaka Sugawara, and Takao Suyama for discussions. 
I also thank the organizers of EAJS especially Hiroshi Itoyama for let me present this work at Osaka.

\end{document}